\documentstyle[preprint,aps]{revtex}
\begin{document}
\draft
\title{\bf Focusing of timelike worldsheets in a theory of strings 
}
\author{Sayan Kar \thanks{Electronic Address :
sayan@iucaa.ernet.in} \footnote{ Present Address : IUCAA, 
Post Bag 4, Ganeshkhind, Pune 411 007, India
}}
\address{{International Centre for Theoretical Physics,
\\ P.O. Box 586, 34100, Trieste, Italy}
\\ and \\ Institute of Physics\\
Sachivalaya Marg, Bhubaneswar--751005, India}
%\twocolumn[
\maketitle
%\widetext
%\parshape=1 0.75in 5.5in
\begin{abstract}
An analysis of the generalised Raychaudhuri equations
 for string world sheets 
is shown to lead to 
the notion of {\em focusing} of timelike worldsheets in the 
classical Nambu--Goto
theory of strings. The conditions under which such
effects can occur are obtained . Explicit solutions
as well as the Cauchy initial
value problem  
are discussed. The results closely resemble 
their counterparts in the theory of point particles which
were obtained in the context of the analysis of spacetime
singularities in General Relativity many years ago. 
\end{abstract}
%\vspace{.3in}
\vskip 0.125 in
\parshape=1 0.75in 5.5in
PACS number(s) : 11.27. +d, 11.10.Kk 
%]

%\narrowtext
\newpage

{\section {\bf Introduction} }
The theory of extended objects such as strings and higher branes
embedded in an ambient background spacetime has been
extensively studied in the recent past. The string/membrane viewpoint has 
found useful applications in  seemingly diverse fields ranging from
a theory of fundamental strings {\cite{ref1:a}} in the context of
quantum gravity and unification
 to two dimensional objects (hypersurfaces) embedded
in an Euclidean background, examples of which are abundant in the
active area of
biological(amphiphilic) membranes {\cite{ref2:b}}. 
Consequences of the generalisation of those equations which
describe various features of a point particle theory to the
case of strings and membranes are therefore worth investigating.

To get into the relevant context we must first ask what these
equations are. For any theory, the starting point is almost always the action. 
The action for the relativistic point particle is the  integral of the 
arc length
$ds$. The first variation results in the equation
of motion which in a general background is the {\em geodesic equation}.
Solutions to this equation are the geodesic curves of the corresponding
background geometry. The second variation of the action 
is related to  the {\em Jacobi/ geodesic deviation } equation   
which governs the separation of one geodesic from another in  a generic
curved background. 
In General Relativity (GR), where spacetime curvature is related to
matter, geodesic deviation provides a measure of the gravitational force
. An alternative set of equations which contain further  information
about the nature
of a one parameter family of geodesics are the {\em Raychaudhuri
equations} {\cite{ref3:c}}. These deal with the issue of
 focusing/defocusing of
geodesic congruences and play a major role in the proofs of the singularity
theorems of GR. 

Each of the above--mentioned equations have generalisations for the 
case of strings as well as higher branes. The geodesic equation is 
replaced by the string/membrane equations of motion and constraints
which emerge out of the first variation of the area functional (Nambu--Goto
action).
The Jacobi equation has also been extended recently by evaluating
the second variation  {\cite{guv:prd}},
{\cite{lf:npb}} (see also {\cite{law:book}} for an earlier reference in the
mathematics literature) . Finally, 
generalised Raychaudhuri equations also exist today due to 
the efforts of Capovilla and Guven {\cite{ref4:d}
. However, not much attention has been
devoted towards understanding the general features of
the solutions of the Jacobi and
Raychaudhuri equations in string/membrane theories in a way
similar to their treatment in the context of GR.  
Our main aim in this paper would therefore be  to analyse the 
 Raychaudhuri 
equation for strings and
derive the worldsheet analog of geodesic focusing .

{\section {\bf The Raychaudhuri equations} }
In introducing the Raychaudhuri equations and its generalisations we shall
prefer writing down the 
equations first and then explain the relevance
and geometrical meaning of the various quantities which appear. 

For the case of families of timelike 
geodesic curves the Raychaudhuri equation
for the quantity known as the expansion $\theta$ is given as :

\begin{equation}
\frac{d\theta}{d\lambda} + \frac{1}{3} \theta^{2} +2 \sigma^{2} -2 \omega^{2}
= - R_{\mu\nu}{\xi}^{\mu}{\xi}^{\nu}
\end{equation}

The expansion $\theta$ measures the rate of change of the cross sectional
area of a family of geodesics. $\sigma^{\mu\nu}$ and 
$\omega^{\mu\nu}$ are known as the shear and 
rotation of the congruence. Thus, if the expansion is negative (positive)  
somewhere we can conclude that the congruence/family is converging (diverging)
. Moreover, if the expansion goes to $-\infty$ we have focusing of geodesics 
-- a generalisation of which is the main topic here.

An alternative way to look at (1) is to convert it into
a second order, linear, ordinary differential equation. 
This is done by a  simple change of variables $\theta = \frac{3}{F}
\frac{dF}{d\lambda}$ which yields the following equation :

\begin{equation}
\frac{d^{2}F}{d\lambda^{2}} + \frac{1}{3}H(\lambda) F = 0
\end{equation}

where $H(\lambda) = R_{\mu\nu}{\xi}^{\mu}{\xi}^{\nu} +2 \sigma^{2} -2\omega^{2}$
The focusing theorem which originates from an analysis of either
version of the Raychaudhuri equation states that
if $\omega^{2} = 0$ and 
matter satisfies an Energy Condition (usually $R_{\mu\nu}{\xi}^{\mu}
{\xi}^{\nu} \ge 0$ or, using Einstein's field equation 
$\left (T_{\mu\nu} - \frac{1}{2}Tg_{\mu\nu}\right ) \xi^{\mu}\xi^{\nu}\ge 0$
) then converging ($\theta $ negative)
families of timelike or null geodesics must necessarily
focus within a finite value of the affine parameter $\lambda$.
Note that the existence of zeros in the class of solutions 
 of (2) implies the divergence of the expansion. 
Detailed analysis of the focusing theorems can be found in 
{\cite{tip:aop77}}, {\cite{wald:85}} and {\cite{he:cup72}}.
Physically, focusing is a natural consequence of the 
{\em attractive} nature of gravitating matter and acts as a pointer
to the existence of spacetime singularities.
 
A generalisation of the above equation is achieved by considering 
families of surfaces as opposed to families of curves. These surfaces
are timelike (i.e. they have a Lorentzian induced metric on the 
worldsheet)  and extremal with respect to variations of the 
Nambu--Goto action. The original derivation for the most
general case of $D$ dimensional timelike, extremal, Nambu--Goto
 surfaces embedded in an
$N$ dimensional Lorentzian background is due to Capovilla and Guven {\cite{ref4:d}}.
The form of the equation 
 for string world sheets given below {\cite{sk:prd96}}
is obtained by using certain properties of two dimensional
surfaces (the choice of isothermal coordinates) and simplifications
achieved by implementing the Gauss--Codazzi integrability conditions.
We have,

\begin{equation}
-\frac{\partial^{2}F}{\partial \tau^{2}} + \frac{\partial^{2}F}{\partial
 \sigma^{2}} + \frac{1}{N-2}\Omega^{2}\left ( -^{2}R + R_{\mu\nu}E^{\mu}_{a}E^{\nu a}
\right ) F = 0
\end{equation}

where $\sigma, \tau$ are the worldsheet coordinates, $^{2}R$ is the
worldsheet Ricci curvature, $R_{\mu\nu}$ is the spacetime Ricci tensor
and $E^{\mu}_{a}$ are the tangent vectors to the worldsheet in the
frame basis ($ g(E_{a}, E_{b}) = \eta_{ab}$ ). $\Omega^{2}(\sigma, \tau)
$ is the conformal factor in the metric induced on the worldsheet
from the background geometry .
We shall denote the coefficient of $F$ in the third term collectively
as $\alpha (\sigma, \tau ) = \frac{1}{N-2}\Omega^{2}\left (
-^{2}R + R_{\mu\nu}E^{\mu}_{a}E^{\nu a} \right )$.

The above generalised equation is a second order, linear, hyperbolic partial
differential equation. It is the parallel of Eqn (2) . We now have 
two quantities $\theta_{\tau}$ and $\theta_{\sigma}$  
which represent the generalised expansions along the $\tau$ and $\sigma$ directions of the
worldsheet and are obtained by taking the partial
 derivative of $\ln F$ with respect to
the $\tau$ and $\sigma$ variables respectively. Our objective now is to 
obtain and analyse
 the solutions of this equation.

{\section {\bf Solutions in light--cone coordinates} } 
In order to arrive at and extract information about the solutions 
of
(3) it is 
useful to make a few assumptions about the quantity $\alpha (\sigma,
\tau)$.
 We can think of two possibilities straightaway. The first of these
is to assume that $\alpha$ is separable in the $\sigma,\tau$ variables
. On the other hand one may prefer going over to light cone coordinates
and assume separability in that system. The conclusions
related to the former case has already been discussed
 in a previous paper by this author {\cite{sk:prd96}}). We therefore
concentrate on the latter.

In light--cone coordinates
defined by :

\begin{equation}
\sigma_{+} = \frac{1}{2}
(\sigma - \tau ) \quad ; \quad \sigma_{-} = \frac{1}{2} (\sigma + \tau )
\end{equation}

the generalised Raychaudhuri equation takes the 
form :

\begin{equation}
\frac{\partial^{2}F}{\partial \sigma_{+}\partial \sigma_{-}} + \
\alpha (\sigma_{+}, \sigma_{-}) F = 0
\end{equation}

where $F$  and $\alpha$ are functions of the $\sigma_{+}, \sigma_{-}$ variables.
A class of solutions of this equation can be easily obtained by
inspection.
We first note that the usual general solution of the wave equation
in $1+1$ dimensions  which involves the superposition
of functions of $\sigma_{+}$ and $\sigma_{-}$ does not work here because 
of the presence of the second term in the equation.

Assuming $\alpha (\sigma_{+}, \sigma_{-}) = \alpha_{+}(\sigma_{+})
\times \alpha_{-}(\sigma_{-})$ we may choose-- 

\begin{equation}
F(\sigma_{+}, \sigma_{-}) = \exp \left ( a\int \alpha_{+}(\sigma_{+})d\sigma_{+}+ b \int \alpha_{-}(\sigma_{-}) d\sigma_{-} \right )
\end{equation}

where $a$, $b$ are two constants which must satisfy the
condition $ab + 1 = 0$ if (6) has to be a solution of (5). 

It is easily seen that the following four possibilities exist for choices
of $a$ and $b$.

\begin{equation}
(1) \quad  a = 1, b = -1 ; \quad (2) \quad a =-1 , b = 1
\end{equation}
\begin{equation}
(3) \quad  a = i, b = i ; \quad (4) \quad  a = -i , b = -i
\end{equation}

Note in the above that there are both oscillatory as well as exponential 
solutions. For the former, we need to look into the real and imaginary 
parts (the cosine and sine solutions respectively) which are --  

\begin{eqnarray}
F(\sigma_{+}, \sigma_{-}) = \cos \left ( \int \alpha_{+}(\sigma_{+})d\sigma_{+} +
\int \alpha_{-}(\sigma_{-})d\sigma_{-} \right )
\\
F(\sigma_{+}, \sigma_{-} ) = \sin \left (\int \alpha_{+} (\sigma_{+})d\sigma_{+}
 +
\int \alpha_{-}(\sigma_{-})d\sigma_{-} \right )
\end{eqnarray}

Introduce the quantities $\theta_{+}$ and $\theta_{-}$ (expansions
along the light cone directions $\sigma_{+}$ and $\sigma_{-}$)
 which are 
related to $\theta_{\sigma}$ and $\theta_{\tau}$ as follows : 

\begin{equation}
\theta_{+} = \theta_{\sigma} -\theta_{\tau} \quad ; \quad \theta_{-} =
\theta_{\sigma} + \theta_{\tau}
\end{equation}

For the exponential solutions we therefore have:

\begin{eqnarray}
\theta_{+} = \frac{1}{F}\frac{\partial F}{\partial \sigma_{+}} = \pm \alpha
_{+}(\sigma_{+}) \\
\theta_{-} = \frac{1}{F}\frac{\partial F}{\partial \sigma_{-}} = \mp \alpha
_{-}(\sigma_{-}) \\
\theta_{+}\theta_{-} = \theta_{\sigma}^{2} - \theta_{\tau}^{2} =
- \alpha (\sigma_{+}, \sigma_{-})
\end{eqnarray}

The upper and lower signs refer to the choices (1) and (2) in Eqn (7)
respectively.

For positive $\alpha$ (i.e. (a) $\alpha_{\pm} > 0$ or (b) $\alpha_{\pm} < 0$
) one can have the following alternatives (we take the lower
sign in the previous expressions, i.e. (2) in Eqn (7))
-- $\theta_{+}$ negative and $\theta
_{-}$ positive (for (a)) and $\theta_{+}$ positive and $\theta_{-}$ negative
(for (b)). On the other hand, for negative $\alpha$ (i.e. (c) $\alpha_{+} > 0$
, $\alpha_{-} < 0$ or (d) $\alpha_{+} < 0$, $\alpha_{-} > 0$ ) the following
possibilities exist-- $\theta_{\pm}$ negative (for (c)) and $\theta_{\pm}$
positive for (d). 
Additionally, 
for this class of solutions, a divergence in $\alpha_{+}$ or $\alpha_{-}$
is necessary to have  divergent expansions. This implies a divergence in 
worldsheet curvature or the spacetime Ricci tensor when evaluated on the 
worldsheet.

Let us now turn to the oscillatory solutions.
The expressions for $\theta_{+}$ and $\theta_{-}$ for them 
can be obtained in a similar fashion. We choose to work 
with the cosine solution  for which 
we have :

\begin{eqnarray}
\theta_{+} = - \alpha_{+}(\sigma_{+}) \tan \left ({\int \alpha_{+}(\sigma_{+}) 
d\sigma_{+} + \int \alpha_{-}(\sigma_{-}) d\sigma_{-}}\right )\\
\theta_{-} = - \alpha_{-} (\sigma_{+}) \tan \left ({\int \alpha_{-}(\sigma_{+})
d\sigma_{+} + \int \alpha_{-} (\sigma_{-}) d\sigma_{-}} \right )\\
\theta_{+}\theta_{-} = \theta_{\tau}^{2} - \theta_{\sigma}^{2} = 
\alpha (\sigma_{+}, \sigma_{-})
\tan ^{2} \left ({\int \alpha_{-}(\sigma_{+}) d\sigma_{+} +
 \int \alpha_{-} (\sigma_{-}) d\sigma_{-}}\right )
\end{eqnarray}

First let us assume $\alpha > 0$ which implies the constraints (a) and (b)
mentioned before on $\alpha_{\pm}$.
Consequently we have $\theta_{\pm}> 0$ or $\theta_{\pm} < 0$ depending on the
sign of the tangent function. On the contrary, if $\alpha < 0$ i.e. cases
(c) and (d), we find that for both cases $\theta_{+}$ and $\theta_{-}$ can only
 have opposite signs.
 However,
in contrast to the oscillatory solutions $\theta_{\pm}$ can diverge
for finite values of $\sigma_{\pm}$ even though $\alpha$ may be completely
regular there.

If $\alpha = 0$ one has to analyse the solutions
of the ordinary wave equation which are given by the
functions $f(\sigma_{+})$ or $g(\sigma_{-})$ or their linear superposition
. By specialising to exponential or oscillatory cases it is
easy to arrive at focusing effects at least for the latter. Note however,
that solutions to the $\alpha \neq 0$ case may not go over smoothly to
those for $\alpha = 0$. The simplest example of this type of
behaviour can be noted for the ordinary differential equation
for the simple harmonic oscillator which has solutions of the form
$\cos kx$, $\sin kx$, $k$ being the frequency. Putting $k =0$ in the
solutions yields trivial results whereas we know that the differential
equation for $k=0$ has a solution of the form $ax+b$ where $a,b$ are
two arbitrary constants.

We now construct an explicit example of an embedding
which is such that the quantity $\alpha$ is separable in
light--cone coordinates.

The background metric is assumed to be conformally flat
--the line element is taken as :

\begin{equation}
ds^{2} = f(x_{0},x_{1}) \left [ -dx_{0}^{2} + dx_{1}^{2} + dx_{2}^{2}
+ dx_{3}^{2}
\right ]
\end{equation}

An embedding which satisfies the Nambu--Goto equations and constraints
could be :

\begin{equation}
x_{0} = C_{1}\tau + C_{2}\sigma \quad ; \quad x_{1} = C_{2} \tau
+ C_{1}\sigma \quad ; \quad x_{2} = constant \quad ; \quad x_{3}
= constant
\end{equation}

where $C_{1}$, $C_{2}$ are constants with $C_{1}^{2} > C_{2}^{2}$.

One therefore needs to write down the expression for the
quantity $\alpha$ which turns out to be :

\begin{equation}
\alpha = -2\frac{1}{\sqrt f} \partial _{+}\partial_{-} \sqrt f
\end{equation}

Defining the induced metric on the worldsheet as :

\begin{equation}
ds_{I}^{2} = e^{2\rho} \left [ -d\tau^{2} + d\sigma^{2} \right ]
\end{equation}

with $e^{2\rho } = f(C_{1}^{2} - C_{2}^{2} )$ we can convert the
expression above into the following form :

\begin{eqnarray}
\alpha = -2e^{-\rho} \partial_{+} \partial_{-} e^{\rho} \nonumber \\
       = -2 \left ( \partial_{+}\partial_{-} \rho + \partial_{+}\rho
       \partial_{-} \rho \right )
\end{eqnarray}

Choosing a generic form of $\rho = A(\sigma_{+}) + B(\sigma_{-})$ 
we can easily see that it is possible to get
an $\alpha$ which is separable in light--cone coordinates.  
Note that in this entire discussion we have never really
chosen an explicit form for the function $f(x_{0}, x_{1})$
. This is not necessary as is apparent from the calculation.
The separability of $\rho$ which ultimately results in the
separability of $\alpha$ however yields a worldsheet metric 
which is flat ($^{2}R = -2e^{-2\rho}\partial_{+}\partial_{-}\rho$ turns out
be zero ).

Also, if the background geometry had been chosen such that the
conformal factor was associated as a factor with the 
$x_{0}$, $x_{1}$ part of the metric (more precisely,
$ds^{2} = f(x_{0}, x_{1}) (-dx_{0}^{2} + dx_{1}^{2}) + dx_{2}^{2}
+ dx_{3}^{2}$) then the same embedding would have resulted in
an $\alpha$ identically equal to zero. 

\section {\bf Focusing theorem}

We now move on to the more important question of analysing the 
generalised Raychaudhuri equation in string theory from the 
viewpoint of a Cauchy initial value problem. Note that the discussion presented
in the previous sections has been largely aimed at obtaining specific solutions
with the assumption of separability in light cone
variables.

Fortunately, we have several oscillation theorems due to Pagan {\cite{ref12:l}}
 and 
co--workers{\cite{ref13:m}}
 which are essentially tailored to our requirements.
We mention below one such theorem which we shall use subsequently.

{\em Theorem : (Pagan and Stocks 1975)}

Let $F(\sigma_{+},\sigma_{-})$ satisfy the  partial differential
equation :

\begin{equation}
F_{+-} + \alpha(\sigma_{+}, \sigma_{-})F = 0
\end{equation}

with the initial conditions :

\begin{equation}
F(\sigma_{+}, \sigma_{+}) = r(\sigma_{+}) \quad ; \quad
\frac{\partial F}{\partial \sigma_{-}} \vert_{\sigma_{+}= \sigma_{-}}
= t(\sigma_{+})
\end{equation}

in the domain $\sigma_{-} -\sigma_{+} \ge  0$ (i.e. $\tau \ge  0)$. Let the
following conditions also hold :

\begin{equation}
\quad (i) \quad \alpha(\sigma_{+}, \sigma_{-})\ge k^{2}> 0 \\
\end{equation}
\begin{eqnarray}
\quad (ii) \quad \alpha_{+} (\sigma_{+}, \sigma_{-}) \ge 0 \quad ; \quad 
\quad (iii) \quad  \alpha_{-} (\sigma_{+}, \sigma_{-}) \ge 0 \\
\quad (iv) \quad  \vert F^{2}(\sigma_{+}, \sigma_{+})- \frac{F_{+}^{2}
(\sigma_{+},
\sigma_{+})}{\alpha (\sigma_{+}, \sigma_{+})} \vert \quad ; \quad 
\vert F^{2} (\sigma_{-},\sigma_{-} ) - \frac{F_{-}^{2}(\sigma_{-}, \sigma_{-}
)}{\alpha (\sigma_{-}, \sigma_{-})} \vert
\\ \quad are \quad bounded \quad 
 as \quad  \sigma_{\pm} \nonumber 
\rightarrow \infty    
\end{eqnarray}

then $F$ changes sign (i.e. develops a zero, nodal line )
 somewhere in the domain :

\begin{equation}
D \equiv \left \{ \sigma_{+}, \sigma_{-} \vert \Sigma_{-}\le \sigma_{-} < 
\infty \quad ; \quad
\Sigma_{+} \le \sigma_{+} < \infty \quad ; \quad \Sigma_{-} - \Sigma_{+}
\ge 0 \right \}  
\end{equation}

The possible existence of the nodal line (a curve along which
$F$ is zero) is the basic result of the abovestated
theorem.  In the language of GR the nodal line is a generalisation
of the focal point-- we might call it the {\em focal curve}
along which families of timelike world--sheets intersect.
 We can see straightaway that there are
several conditions which have to be obeyed in order to ensure the
existence of a nodal line. We  now briefly discuss the
implications of each of them. 

The condition (i) is the analog of the usual 
Energy Condition in the theory of geodesic curves although the 
R. H. S. of the inequality has a {\em positive number} instead of 
zero. However, since the $\alpha = 0$ case leads to a
simple wave equation (whose solutions always have
zeros) we can extend this condition to the
analog of the usual energy condition with the $k^{2}$ being
replaced by zero.

The second condition (i.e. (ii)) imposes restrictions on the
derivatives of $\alpha$. Translated in the language of 
$\sigma, \tau$ coordinates one can easily check that
the following have to hold true.

\begin{equation}
\frac{\partial \alpha}{\partial \sigma} \ge \frac{\partial \alpha}
{\partial \tau} \quad ; \quad \frac{\partial \alpha}{\partial \sigma}
\ge - \frac{\partial \alpha}{\partial \tau}
\end{equation}

Therefore, if $\alpha$ is only a function of $\sigma$ one 
requires $\frac{d\alpha}{d\sigma} \ge 0$ whereas if
$\alpha$ is only a function of $\tau$ then one actually
ends up in a contradiction , the only resolution of
which is to assume $\alpha$ as a constant or zero.

Finally, the third condition, which is on the function $F$
, implies, for instance, for the oscillatory solutions
derived earlier, the boundedness of the quantity $\alpha$ as $\sigma_{\pm}$
approaches $\pm \infty$. This can be seen by substituting the
solution in the expression for the condition. 

Based on the above theorem we can now frame our focusing theorem for 
timelike worldsheets.

{\em If $\theta_{+}$ or $\theta_{-}$ is negative somewhere then 
they tend to $-\infty$ within a finite value of the worldsheet 
parameters $\sigma_{+}$ or $\sigma_{-}$ provided all
conditions on the $\alpha$ are obeyed. The negativity of $\theta_{+}
$ or $\theta_{-}$ is dependent on the negativity of the functions
$r$ and $t$ which appear in the initial conditions.}

 It is perhaps easier to visualise the notion of focusing for the case
of a family of closed string worldsheets. Assume a family
of cylindrical worldsheets which meet along some curve
$\sigma_{+} = f(\sigma_{-})$. This curve is the nodal line
mentioned before.  It may happen, that this 
curve (nodal line) degenerates to a point. 
For example, if the equation of the curve turns out to be 
$\sigma_{+}^{2} + \sigma_{-}^{2} = 0$. Then the only
real solution is $\sigma_{+}= \sigma_{-} =0$. For such cases 
we have a family of cones
emerging out of that point--the common vertex of the cones being the
focal point of the congruence of worldsheets. This basically means
that the worldsheet geometries have a conical singularity in the sense
of unbounded curvature at that specific point. In GR, the focal point
of a congruence indicates a singularity in the congruence of geodesics.
We may find that the spacetime singularity (in the sense
of unbounded curvature) coincides with the focal point of a 
geodesic congruence
in the spacetime--
for example, this happens in the universe models which exhibit 
curvature singularities
. However, this is not true always.
At this stage, it is not completely clear whether a notion of incompleteness of
 string worldsheets
 can be derived and related to a singularity in the background spacetime.
Moreover, we do not know precisely if the conical worldsheet singularity
which may arise if the focal curve degenerates to a point
has any relation with the background spacetime singularities. 
Further analysis is essential if one wishes to arrive at a 
better understanding.

It must be mentioned that this is $a$ focusing theorem -- i.e. with
the assumptions on the various quantities one can conclude that a
focal curve  can exist. The results of the previous section are different
from two angles--firstly we do {\em not} frame an initial value problem
there and secondly the solutions are obtained ad--hoc, largely by
inspection. It is possible that under different assumptions on the
variables as well as other initial conditions one may also be able to 
prove the existence of a nodal curve.
The author, unfortunately, is unaware of such results either in the mathematics
or in the physics literature. 

{\section {\bf Offshoots} }
Before we conclude let us point out certain applications of the 
formalism of the generalised Jacobi and Raychaudhuri equations
in a completely different context--that of biological membranes
. Here, we consider two--dimensional hypersurfaces embedded in an
Euclidean (flat) background space. The first variation
yields the surface configurations which can be minimal
(zero mean curvature)  or Willmore (constant mean
curvature) depending on the choice of the action
functional.  The corresponding Jacobi equations  
contain information about the normal deformations of these two--dimensional
surfaces and are thereby linked to the question of stability. 
 Much of the basic notions along
these directions have been pursued in a mathematical context
in a number of papers {\cite{tuz:93}}. However, an application 
oriented analysis with an emphasis on specific, physically relevant cases 
has not yet been performed.
On the other front, solutions of the Raychaudhuri equations, which are 
in a certain sense nonperturbative, would indicate  the formation of
cusps and kinks on the membrane (focusing along a degenerate
curve(point)). 
A major difference with the 
analysis presented in this paper and that required to
understand membranes in an Euclidean background is the appearance
of elliptic equations as opposed
to hyperbolic ones. Therefore, to analyse the Jacobi equations/ focusing one 
has to
utilise the oscillation theorems for elliptic equations. Fortunately,
once again such theorems do exist {\cite
{kr:ell}}. 
A detailed presentation of these ideas and their consequences in the
context of biological membranes will be reported elsewhere {\cite{sk:prep}}.
 
{\section {\bf Conclusions} }
In conclusion we summarise and raise a few questions of related interest.

We have obtained a {\em focusing theorem} for string worldsheets. This
is illustrated through exact solutions as well as an analysis of the
Cauchy initial value problem. The conditions for focusing are
outlined--these constrain worldsheet as well as spacetime properties.
An analysis of the Jacobi equation  as well as a more detailed 
presentation of the ideas here is in progress and will be 
reported in the near future {\cite{sk:prep1}}.

It is a somewhat pleasing fact that most of the results for point
particle theories have their generalisation for the case of strings.
However, GR as a theory of gravity has an unique feature--the equations
of motion for test particles (i.e. the geodesic equation)
 can be derived from the
Einstein field equations for the field $g_{\mu\nu}$
{\cite{eins:ref}}.
We may therefore ask--given the string equation of motion can one find 
the corresponding `Einstein equation'
which would lead to it under the suitable assumptions which may 
define a {\em test} string?

Finally, of course, one has to address the question of background spacetime
 singularities 
-- does 
a string description as opposed to the point particle resolve the
issue at the classical level ? As a first step towards this
(following the path of GR) we now have a focusing theorem.
It would perhaps be worthwhile to attempt a derivation of the analogs of the 
Hawking--Penrose theorems for the case of strings and thereby
demonstrate the existence/non--existence of spacetime singularities in
the classical theory of strings. If the answer remains the same as in GR (i.e. spacetime
singularities exist under quite general conditions) then one 
can proceed towards examining how quantum string theory can 
help us solve the problem.

The author wishes to thank S. Randjbar--Daemi for inviting him to
visit the International Centre for Theoretical Physics, Trieste,
Italy where this work was done.

\end{document}